\documentclass[sort&compress
   ,final            % use final for the camera ready runs
]{aipproc}
\layoutstyle{8x11double}
\usepackage{color}
\usepackage{nicefrac}

\usepackage{here}
\usepackage{array}
\usepackage{booktabs}
\begin{document}

\title{Revealing Differences Between Curricula Using\\the Colorado Upper-Division Electrostatics Diagnostic}



\author{Justyna P. Zwolak}{address={Department of Physics, Oregon State University, 301 Weniger Hall, Corvallis, OR 97331}}

\author{Corinne A. Manogue}{address={Department of Physics, Oregon State University, 301 Weniger Hall, Corvallis, OR 97331}}

%%%%%%%%%%%%%%%%%%%%%%%%%%%%%%%%%%%%%%%%%%%%
\begin{abstract}
The Colorado Upper-Division Electrostatics (CUE) Diagnostic is an exam developed as part of the curriculum reform at the University of Colorado, Boulder (CU). It was designed to assess conceptual learning within upper-division electricity and magnetism (E\&M). Using the CUE, we have been documenting students' understanding of E\&M at Oregon State University (OSU) over a period of 5 years. Our analysis indicates that the CUE identifies concepts that are generally difficult for students, regardless of the curriculum. The overall pattern of OSU students' scores reproduces the pattern reported by Chasteen et al. at CU. There are, however, some important differences that we will address. In particular, our students struggle with the CUE problems involving  separation of variables and boundary conditions. We will discuss the possible causes for this, as well as steps that may rectify the situation. 
\end{abstract}

\maketitle

%%%%%%%%%%%%%%%%%%%%%%%%%%%%%%%%%%%%%%%%%%%%
%% MAINMATTER
%%%%%%%%%%%%%%%%%%%%%%%%%%%%%%%%%%%%%%%%%%%%
\section{Introduction}

\hspace{0.35cm}Standardized assessment tests that allow researchers to compare the performance of students taught according to various curricula are highly desirable. Such comparisons provide information about the effectiveness of different curricula and, as a result, can improve methods of teaching, learning trajectories and, ultimately, student learning. Appropriately designed diagnostics not only reveals common student difficulties but can also help to determine to what extent students understand the content. 

As of the present day, there are several research-based conceptual tests that serve as instruments to assess and identify students' difficulties in lower-division courses (e.g., the Force Concept Inventory \cite{Hestenes92-FCI}, the Conceptual Survey of E\&M \cite{Maloney01-CSEM},  the Basic Electricity and Magnetism Survey \cite{Ding06-BEMA}). Data from these tests help to determine, among other things, where students lack a conceptual understanding of the material and help to correlate this with various methods of teaching. It also allows teachers and researchers to find out if these difficulties are present more universally.

At the upper-division level, assessing students' difficulties is a much more challenging task. Several research groups are currently working on such tests (e.g., the Colorado Upper-Division Electrostatics \cite{Chasteen09-CUE1,Chasteen12-CUE,Pepper12-SDM}, Colorado UppeR-division ElectrodyNamics Test \cite{Baily12-AUE}, the Quantum Mechanics Assessment Tool \cite{Goldhaber09-TQM}, the Survey of Quantum Mechanics Concepts \cite{Singh05-AQM}). These upper-division assessments are relatively new and thus they have only been employed at a few institutions. 

In the Paradigms in Physics program at OSU, we instituted a radical reform of all the upper-division physics courses that led to extensive reordering of the content. In traditional curricula, courses focus on a particular subfield of physics (e.g., classical mechanics, electricity and magnetism, quantum mechanics). At OSU, courses -- called Paradigms -- revolve around concepts underlying those fields (e.g., energy, symmetry, forces, wave motion). Therefore, the content is arranged differently -- in E\&M we spend more time on direct integration and curvilinear coordinates, less time on separation of variables, we cover potentials before electric fields and magnetostatics in vacuum before electrostatics in matter. The gravitational analogue of electrostatics is covered at the same time as electrostatics rather than in a classical mechanics course. Moreover, we use a large variety of active engagement strategies, such as individual small whiteboard questions, small group problem-solving, computer visualization and kinesthetic activities \cite{OSU-activ}. 

The Paradigms courses, taken in the junior year, are then followed by Capstone courses, which have a more traditional, lecture-based structure. Our students thus represent an important test case to examine the versatility of new assessment tools. In this paper, we will focus on the Colorado Upper-Division Electrostatics diagnostic \cite{Chasteen09-CUE1,Chasteen12-CUE,Pepper12-SDM}. We have found significant value in this new assessment tool. Using the CUE in our own classes has already pointed out several possibilities for the improvement of the curriculum at OSU. However, this new measure is still in the developmental stage and requires some fine tuning when used outside the University of Colorado. Our goal is to help generalize the CUE to be accessible and relevant at a range of institutions. 

\begin{table}[t]
\renewcommand{\arraystretch}{1.2}
\begin{tabular}{c c c || c}
\cline{1-4} 
\multicolumn{3}{c||}{Junior Courses} & Senior Courses \\ 
Fall & Winter & Spring & Fall \\ \cline{1-4}
\rule{0pt}{25pt} \parbox[c]{3cm}{\centering \textbf{Symmetries}\\ \textbf{Vector Fields}\\ Oscillations} &\parbox[c]{3cm}{\centering Preface\\Spins\\\emph{1-D Waves}\\\emph{Central Forces}\vspace{3pt}} & \parbox[c]{3cm}{\centering Energy and Entropy\\Periodic Systems\\Reference Frames\\Classical Mechanics\vspace{2pt}} & \parbox[c]{3cm}{\centering \emph{Mathematical Methods}\\\textbf{Electromagnetism}} \\ 
\cline{1-4}
\end{tabular}
\caption{Standard schedule of Paradigms and Fall term Capstones (for full schedule see Ref. \cite{PinP}). E\&M-related courses, during which the CUE is being administered, are highlighted in bold, courses where the method of separation of variables is discussed are in italics.}
\label{tab:course_desc}
\end{table}

%%%%%%%%%%%%%%%%%%%%%%%%%%%%%%%%%%%%%%%%%%%%
\section{Methodology}
	
\hspace{0.35cm}The  CUE was originally designed as a free-response conceptual survey of electrostatics (and some magnetostatics) for the first semester of an upper-division level E\&M sequence. It is designed in a pre/post format. The 20-minute pre-test contains 7 out of the 17 post CUE questions that junior-level students might reasonably be expected to  solve based on their introductory course experience. The post-test is intended to be given at the end of the first upper-division semester in a single 50-minute lecture. Instead of actually solving problems, students are asked to explain how they would solve them. They are rated for both choosing the appropriate method and the correctness of their reasoning in deciding on a given method. The instructions for the first half of the post-test are as follows:
\begin{quote}
For each of the following, give a brief outline of the EASIEST method that you would use to solve the problem. Methods used in this class include but are not limited to: Direct Integration, Ampere's Law, Superposition, Gauss' Law, Method of Images, Separation of Variables, and Multipole Expansion.
\end{quote}

\begin{figure}[b]
 \includegraphics[width=0.48\textwidth]{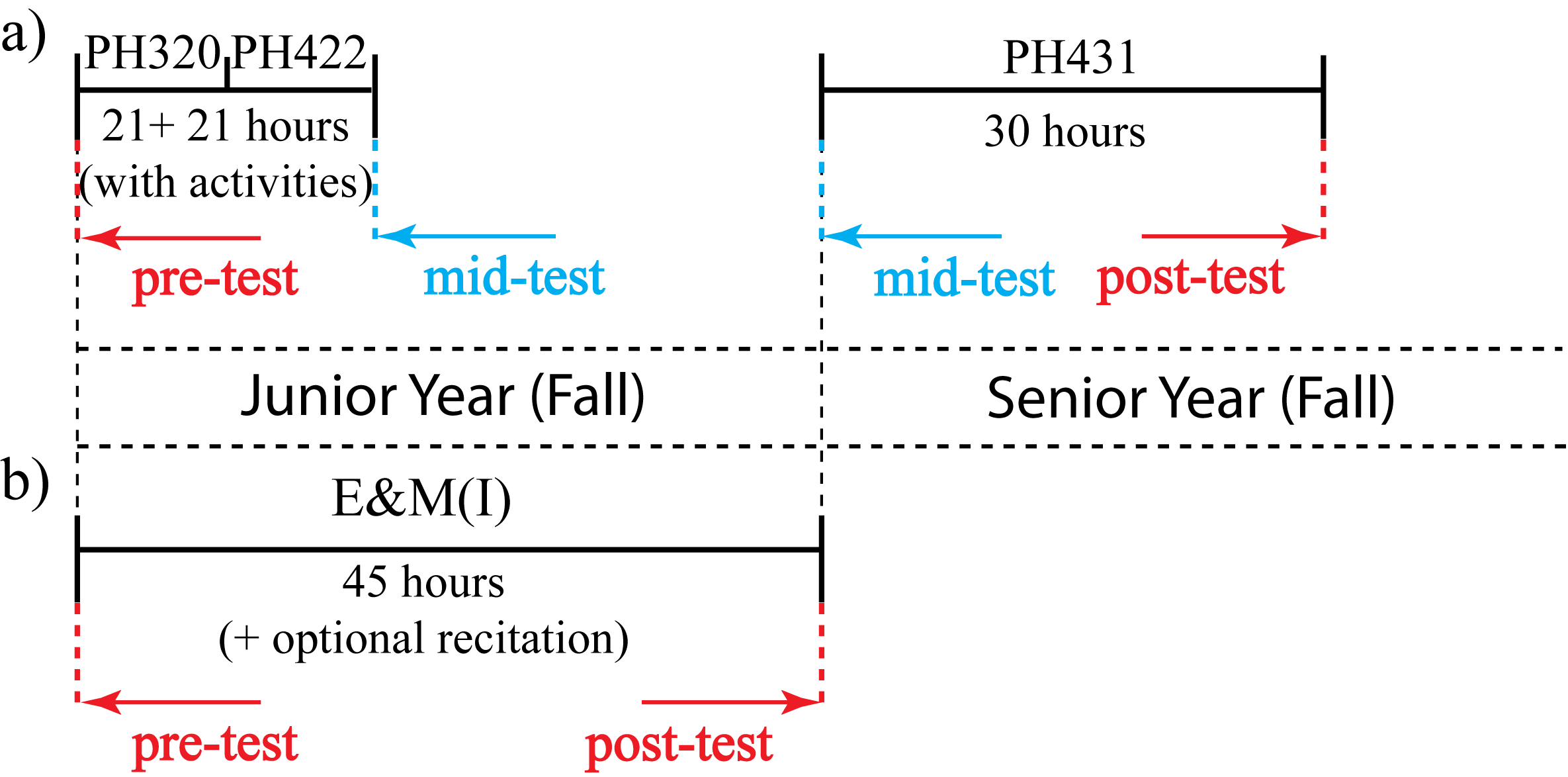}
 \caption{Schedule of administering the CUE at (a) OSU (quarter systems) and (b) CU (semester system). The timescale here shows that the CU E\&M(I) course occurs over 15 weeks whereas the PH320 and PH422 Paradigms are more intense and last 3 weeks each.}
\label{fig:emscheme}
\end{figure}

\begin{figure}[!ht]
\includegraphics[width=0.7\textwidth]{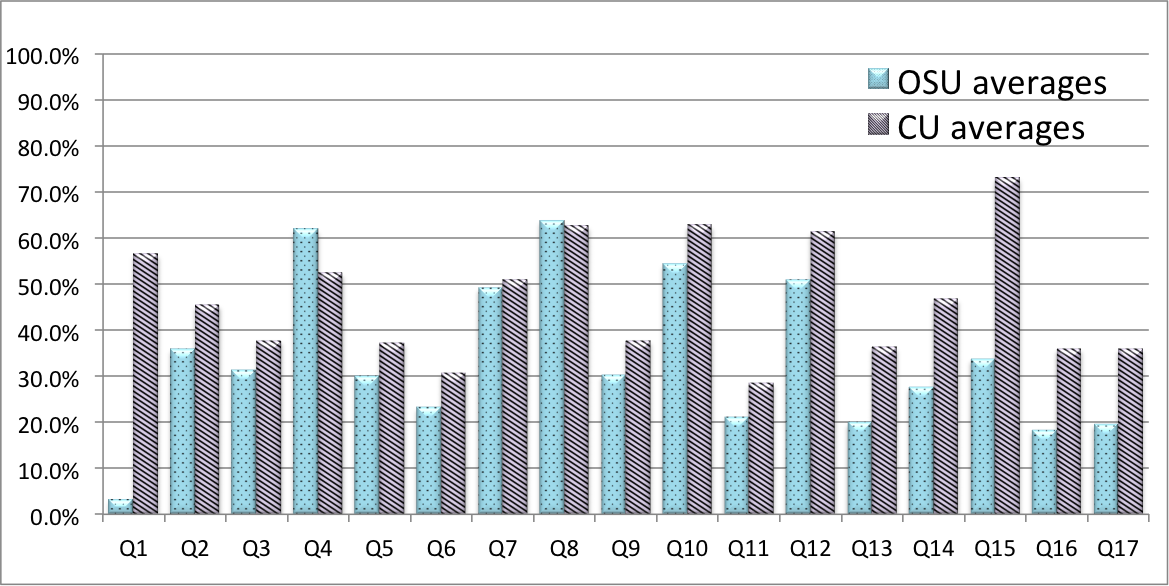}
\caption{Mean values for each question on the CUE for OSU ($N=37$, blue dotted pattern) and for CU ($N=103$, purple hatched pattern).}
\label{fig:post_total}
\end{figure}

We have been using the CUE to assess students' understanding of junior-level E\&M over a period of 5 years (from 2009 to 2013). At the beginning of the Fall term of each year, junior-level students enrolled in Symmetries and Idealizations Paradigm (PH320) took the full CUE pre-test (see TABLE~\ref{tab:course_desc} for the OSU course schedule and FIGURE~\ref{fig:emscheme} for a timeline of the CUE at OSU and CU). The same group of students was given the mid-test (a subset of 12 post-test questions we chose to conform to our course goals) at the end of the Static Vector Fields Paradigm (PH422/522). In the following year, those students who successfully finished their junior year were again given two tests within the Electromagnetism Capstone course (PH 431). There was a second mid-test at the beginning of the term (with the same set of 12 questions as in the first mid-test) and the full CUE post-test at the end of the term. 

The necessity of introducing the mid-test arose due to the different course structure at OSU. Since not everything that the CUE tests is covered by the end of the fall quarter, the results from a full CUE post-test would not have been appropriate.  We also note that, although OSU students have had more contact hours in E\&M (72 hours) at the time they take the post-test than CU students (45 hours), most of the additional hours are on the more advanced content from E\&M(II) at CU.

%%%%%%%%%%%%%%%%%%%%%%%%%%%%%%%%%%%%%%%%%%%%
\section{CUE at Oregon State University}

\hspace{0.35cm}The CUE post-test was administered three times between the Fall term of 2010 and the Fall term of 2013 (with the exclusion of the Fall term of 2012). The timing of the test was consistent throughout that period -- each time it was given in the last week of the term. A total number of $N=39$ students took the CUE post-test, out of whom two students were excluded from the research (one student was a member of the PER group at OSU and participated in meetings where the CUE was discussed, the other student took only some of OSU's E\&M courses and therefore did not take a sequence of at least two tests). 

Students at OSU scored on average $36.5\pm2.4\%$ (compared to $47.8\pm1.9\%$ at CU reported in Ref.~\cite{Chasteen12-CUE}), with the spread of their performance ranging from about $12\%$ to $70\%$. To provide a measure of student improvement over time we used the (non-)normalized gain proposed in Ref.~\cite{Hake97-ANG}. The normalized gain is defined as the ratio of the actual average gain to the maximum possible average gain,
\begin{equation}
\mathbf{g}_{nor}=\frac{\langle 7Q\,post\textrm{-}test\rangle - \langle pre\textrm{-}test\rangle}{100-\langle pre\textrm{-}test\rangle}\,,
\end{equation}
 where by $\langle 7Q\,post\textrm{-}test\rangle$ we denote the average score of a given student from the subset of post-test questions which match the pre-test problems. For the $N=24$ students who took both the pre- and post-tests, we found an average normalized gain of $33\%$ ($28\%$ non-normalized\footnote{Non-normalized (absolute) gain is an actual average gain calculated as $\mathbf{g}_{abs}=\langle 7Q\,post\textrm{-}test\rangle-\langle pre\textrm{-}test\rangle$.}), which is similar to gains of $34\%$ (normalized) and $24\%$ (non-normalized) reported in Ref.~\cite{Chasteen12-CUE}. Thus, although students at OSU on the average scored about $11\%$ lower than students at CU on both the pre- and post-tests\footnote{The average score on the final version of the CUE pre-test was $19.6\pm1.9\%$ at OSU (for $N=76$) and $29.4\pm2.4\%$ at CU (for $N=51$), see Ref.~\cite{Chasteen12-CUE}.}, they showed similar learning \emph{gains} to students from other institutions taught in PER-based courses and higher gain than were observed in standard lecture-based courses (see FIGURE 7 in Ref.~\cite{Chasteen12-CUE}). 

FIGURE ~\ref{fig:post_total} shows comparison of the average performance on each question between students from OSU (blue dotted plot) and CU\footnote{Data adapted from Appendix A: ``Student Performance on Final CUE Questions'', in Ref.~\cite{Chasteen12-CUE}} (purple hatched plot).  One of the most striking features of this plot is the similarity of the overall pattern of students' scores -- both on the high- and low-scored questions.  With the exception of two questions (Q1 and Q15), the averages agree to within $10\%$ on the first 12 questions and to within $20\%$ thereafter. It is also worth noting that, despite the low number of students taking the CUE post-test in individual years, this pattern is still preserved when we plot the average scores of each question by year. This affirms the reliability of the CUE across the two very different curricula. Moreover, the surprisingly low average scores on some questions from both institutions suggest that the CUE is a very challenging test in general, regardless of the curriculum. 

\begin{figure}[t]
 \includegraphics[width=0.48\textwidth]{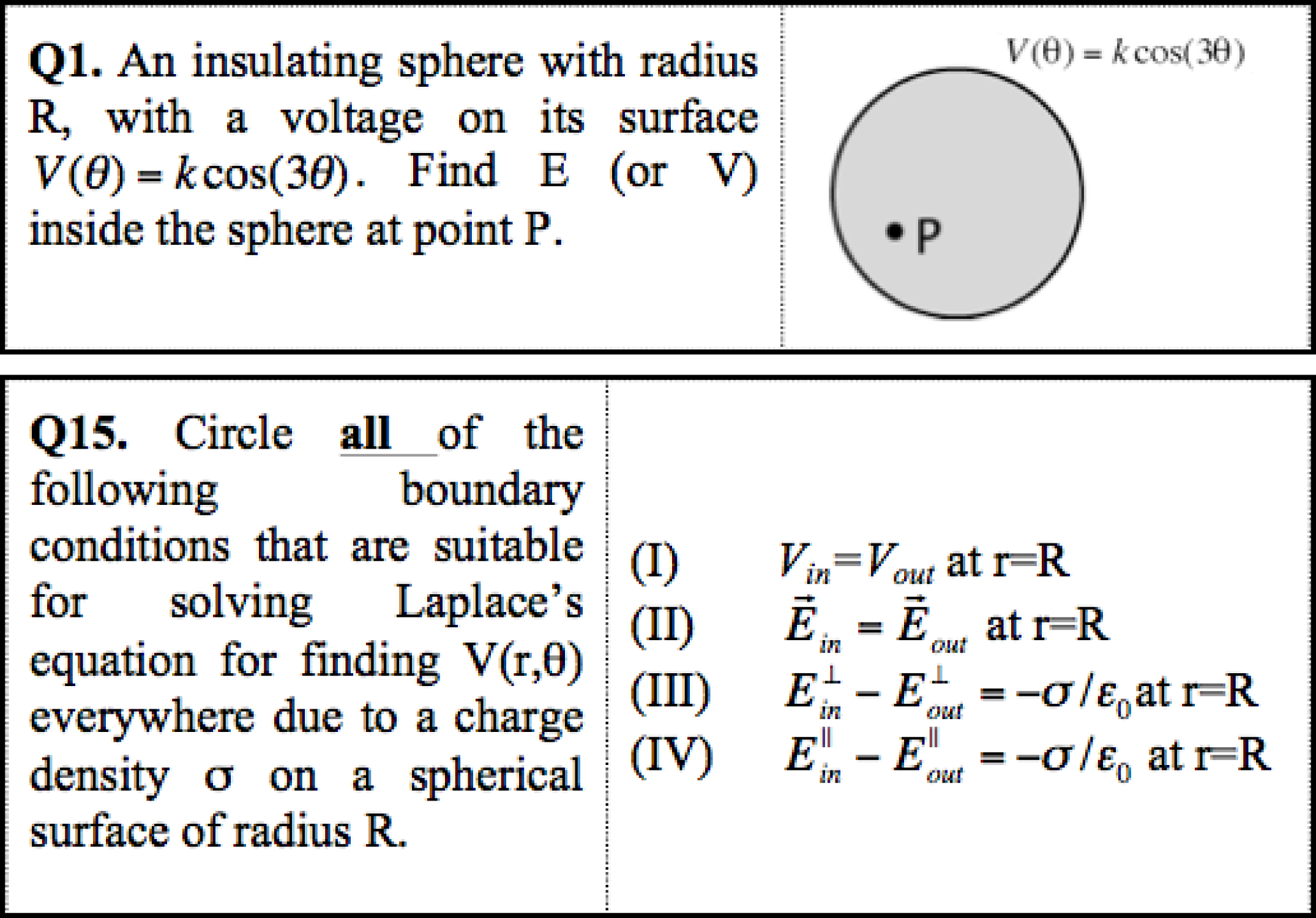} 
\caption{CUE questions where the scores of OSU students differ significantly from the scores of students taught at CU.}
\label{fig:Q1+Q15}
\end{figure}

Although the overall pattern in FIGURE~\ref{fig:post_total} from both institutions is very similar, there are some significant differences that need to be addressed. In particular, OSU students' scores differ by over $50\%$ on question Q1 regarding finding the potential $V$ (or field $E$) inside an insulating sphere and by almost $40\%$ on question Q15 regarding selecting boundary conditions to solve for the potential $V(r,\theta)$ on a charged spherical surface. For reference, the full problems are reproduced in FIGURE ~\ref{fig:Q1+Q15}. Both of these questions are intended to test whether students can set up the solution to a problem involving partial differential equations (i.e, recognizing separation of variables as an appropriate problem-solving technique and/or defining appropriate boundary conditions) \cite{Chasteen12-CUE,CEMLG}. To address these discrepancies we now look more closely at the learning goals for courses at OSU.

In a traditional curriculum, as defined by the standard E\&M text by David Griffiths \cite{Griffiths-EM}, students are first exposed to the application of separation of variables in physics in their E\&M course, before they take quantum mechanics. At OSU, however, students are exposed to the separation of variables mainly in the context of the Schr\"{o}dinger equation -- first in the Waves in One Dimension Paradigm and in the Central Forces Paradigm in the Winter term of the junior year and then in the Mathematical Methods Capstone in the Fall term of the senior year. Separation of variables is being discussed in multiple courses before students take the Capstone in E\&M and thus not much time is devoted to this topic in the E\&M Capstone. To be precise, there is only one day (typically the second day of the first week) spent on Laplace's equation, followed by 2 or 3 homework problems. As a consequence, students have much more experience with separation of variables in the context of quantum mechanics, long before they see it as part of E\&M, and even then the structure of the Capstone does not provide them with many opportunities to practice it in the E\&M context. Low scores on the two other questions involving separation of variables and boundary conditions (BCs): Q11 (finding BCs in a specific scenario) and Q13 (recognizing the form of solutions that match given BCs) supports our suspicion that students are not getting enough exposure to these topics in the context of E\&M.

%%%%%%%%%%%%%%%%%%%%%%%%%%%%%%%%%%%%%%%%%%%%%%%%
\section{Discussion}

\hspace{0.35cm}Due to the significantly restructured curriculum at OSU, our findings provide valuable data for comparison with reported results from CU's more moderately reformed curriculum and from institutions with a more traditional (lecture) format. It is intriguing that the across curricula difficulty pattern for most questions is preserved, even though the sample of students is quite different. This result confirms the overall robustness of the CUE.  In addition, strong differences in scores on a few specific questions shows that this assessment test is capable of helping to distinguish between different programs of study and uncovering important gaps in a curriculum. 

It is crucial to understand the causes for the big discrepancies between OSU and CU scores on Q1 and Q15. As we indicated above, one of the reasons might be the current organization of courses at OSU. While restructuring the junior- and senior-level courses at OSU, it was assumed that -- once exposed to certain techniques of solving problems in one context -- students will be able to transfer their knowledge of its applicability from one subfield of physics to another. As the CUE has revealed, however, this is not happening and the separation of variables procedure does not become a natural E\&M problem-solving technique for students once they depart from the quantum world. OSU has made a recent change in the schedule of the Paradigms and Capstones -- moving the Mathematical Methods Capstone, as well as, the Central Forces Paradigm to the Spring term of the junior year. This rearrangement gives us an opportunity to test whether the inclusion of more examples where separation of variables and boundary conditions are explicitly used to solve E\&M problems can impart the generality of the techniques to the students and subsequently be reflected in higher CUE scores on the relevant questions. We will discuss this change in a later publication.

%%%%%%%%%%%%%%%%%%%%%%%%%%%%%%%%%%%%%%%%%%%%%%%%
%% BACKMATTER
%%%%%%%%%%%%%%%%%%%%%%%%%%%%%%%%%%%%%%%%%%%%%%%%

\begin{theacknowledgments}
Supported in part by NSF DUE 1023120 and 1323800. We would like to thank Steve Pollock and Bethany Wilcox for conversations about the design and grading of the CUE and Stephanie Chasteen for helping us with the CU test data.

\end{theacknowledgments}



\bibliographystyle{aipproc}   % if natbib is available

%\bibliography{EdR}

\IfFileExists{\jobname.bbl}{}
 {\typeout{}
  \typeout{******************************************}
  \typeout{** Please run "bibtex \jobname" to optain}
  \typeout{** the bibliography and then re-run LaTeX}
  \typeout{** twice to fix the references!}
  \typeout{******************************************}
  \typeout{}
 }

\end{document}